\documentclass[12pt,aps,nofootinbib,preprintnumbers]{revtex4}

\usepackage{amsmath,amssymb,bm}
\usepackage[dvips]{graphicx}
\usepackage{yfonts}
\newcommand{\beq}{\begin{eqnarray}}
\newcommand{\eeq}{\end{eqnarray}}

\newcommand{\SU}{\text{SU}}
\newcommand{\U}{\text{U}}
\newcommand{\tr}{\mathop{\mathrm{tr}}}
\newcommand\LambdaQCD{\Lambda_{\rm QCD}}
\newcommand\stru{\rule[-15pt]{0pt}{15pt}}

\begin{document}

\preprint{INT-PUB-10-052}
\author{Dam~T.~Son and Naoki Yamamoto}
\affiliation{Institute for Nuclear Theory, University of Washington, 
Seattle, Washington 98195-1550, USA}

\title{Holography and Anomaly Matching for Resonances}

\begin{abstract}
We derive a universal relation for the transverse part of triangle anomalies
within a class of theories whose gravity dual is described by 
the Yang-Mills-Chern-Simons theory.
This relation provides a set of sum rules involving the masses, decay
constants and couplings between resonances, and leads to the formulas 
for the matrix elements of the vector and axial currents 
in the presence of the soft electromagnetic field.
We also discuss that this relation is valid in real QCD
at least approximately.
This may be regarded as the anomaly matching for resonances
as an analogue of that for the massless excitations in QCD.

\end{abstract}

\maketitle 

\section{Introduction}

One distinctive feature of relativistic quantum field theories 
is the existence of anomalies \cite{Adler:1969gk, Bell:1969ts, Adler:1969er},
which is the violation of some symmetries of the classical action by
quantum effects.
In the case of global symmetries, when currents are coupled to  
external gauge fields, not all currents can be conserved.  This fact is
reflected in the longitudinal part of the triangle diagrams.
The longitudinal part of triangle anomalies 
does not depend on the energy scale due to its topological nature: 
the triangle anomalies calculated in QCD at the level of quarks and gluons 
are reproduced at the level of hadrons 
(the 't Hooft anomaly matching condition) \cite{'tHooft:1980xb};
this leads to observable consequences for the low-energy physics involving 
pions in QCD. A well-known example is the $\pi_0 \rightarrow 2\gamma$ decay.
One can ask if the transverse part of the triangle graphs is also constrained.
If such a constraint exists, it would have implications for the physics of
hadron resonances (the $\rho$ and $a_1$ mesons, in particular).

Such a question was posed in Ref.~\cite{Vainshtein:2002nv} and further
studied in Refs.~\cite{Czarnecki:2002nt,Knecht:2003xy}.  It was found
that the transverse part of the current-current correlator in an
infinitesimally weak electromagnetic field [denoted as $w_T(Q^2)$ and defined
below] is not renormalized in perturbative QCD, and so the transverse
part is related to the longitudinal part. However, chiral symmetry
breaking leads to a violation of this relationship.  The
nonperturbative aspects of the transverse part have been studied
mostly at large Euclidean momentum $Q^2=-q^2$.  Clearly, the main
difficulty is that the transverse part of triangle anomalies has a
dynamical nature rather than a topological one.

In this paper, we study the transverse part of triangle anomalies
using the technique of holography \cite{Maldacena:1997re,
Gubser:1998bc, Witten:1998qj}.  We consider first a class of theories
whose gravity dual is described by the Yang-Mills-Chern-Simons theory
with chiral symmetry broken by boundary conditions in the infrared.
This class of theories include the early ``bottom-up" AdS/QCD model
inspired by dimensional deconstruction and hidden local 
symmetry~\cite{Son:2003et} and the ``top-down" Sakai-Sugimoto model
\cite{Sakai:2004cn}.
(Both models reproduce rather well various aspects of 
the physics of low-lying hadrons in QCD.)
For models in this class, we derive
the following relation for the transverse part of triangle anomalies:
\beq
\label{eq:result}
  w_T(Q^2)=\frac{N_c}{Q^2}-\frac{N_c}{f_{\pi}^2}
  \left[\Pi_{A}(Q^2) - \Pi_{V}(Q^2)\right],
\eeq
for {\it any} $Q^2$. 
Here $w_T(Q^2)$ is defined in Eq.~(\ref{eq:expansion}) below,
$N_c$ is the number of colors, $f_{\pi}$ is the pion decay constant, 
and $\Pi_A$ and $\Pi_V$ are the axial and vector current correlators, 
respectively.
Equation~(\ref{eq:result}) fully includes the nonperturbative correction 
and may be regarded as the ``anomaly matching for resonances" as an 
analogue of that for the massless excitations in QCD.
As will be shown, Eq.~(\ref{eq:result}) provides a set of sum rules 
involving the resonance parameters, 
leading to the formulas for the matrix elements of the vector 
and axial currents in the presence of the soft electromagnetic field
[see Eqs.~(\ref{eq:mat-ele2}) and (\ref{eq:mat-ele3})].

We also argue that Eq.~(\ref{eq:result}) holds 
at least approximately in real QCD at both small and large $Q^2$.

\section{Triangle anomalies}
\label{sec:triangle}
First we review the triangle anomalies. 
We consider massless QCD with $N_c$ colors and $N_f$ flavors.
Let us define the correlation function of the vector current
$j_{\mu}^a=\bar q \gamma_{\mu} t^a q$ and the axial current
$j_{\mu}^{5b}=\bar q \gamma_{\mu} \gamma_5 t^b q$ 
in a weak electromagnetic background field 
$\hat F_{\mu \nu}= \partial_{\mu} \hat V_{\nu}- 
\partial_{\nu}\hat V_{\mu}$,
\beq
  \label{eq:correlation}
  d^{ab} \langle j_{\mu} j_{\nu}^5 \rangle_{\hat F}
  \equiv
  i\!\int\! d^4 x\, e^{iqx}\langle j_{\mu}^a(x)j_{\nu}^{5b}(0) 
  \rangle_{\hat F},
\eeq
where $t^a$ ($a=1,2, \cdots, N_f^2-1$) and $t^0={1}/{\sqrt{2N_f}}$
are the $\U(N_f)$ flavor matrices 
normalized so that $\tr(t^a t^b)=\delta^{ab}/2$. 
We also define $d^{ab}=(1/2)\tr({\cal Q} \{t^a,\, t^b \})$ 
where ${\cal Q}$ is the electric charge matrix.
Since $\langle j_{\mu} j_{\nu}^5 \rangle_{\hat F}$ is a Lorentz pseudo-tensor,
the leading term in its expansion over the weak background field 
$\hat F_{\mu \nu}$ is a linear combination of three structures: 
$\tilde F_{\mu \nu}$, $q_{\mu}q^{\sigma} \tilde F_{\sigma \nu}$, 
and $q_{\nu}q^{\sigma} \tilde F_{\sigma \mu}$ with 
$\tilde F_{\mu \nu} = (1/2)\epsilon_{\mu \nu \alpha \beta} 
\hat F^{\alpha \beta}$.  
Imposing vector current conservation
$q^{\mu} \langle j_{\mu} j_{\nu}^5 \rangle_{\hat F}=0$, the number of
independent structures reduces to two:
the longitudinal and transverse parts with respect to $q^{\nu}$.
The general expression up to the leading order in $\tilde F$ is
\beq
\label{eq:expansion}
\langle j_{\mu} j_{\nu}^5 \rangle_{\hat F} =-\frac{1}{4\pi^2}
\left[w_T(q^2)(-q^2 \tilde F_{\mu \nu} + q_{\mu}q^{\sigma} 
  \tilde F_{\sigma \nu}
-q_{\nu}q^{\sigma} \tilde F_{\sigma \mu}) 
+ w_L(q^2) q_{\nu}q^{\sigma} \tilde F_{\sigma \mu} \right],
\eeq 
where we follow the notation of \cite{Vainshtein:2002nv}.
The longitudinal and transverse nature of the terms in this expression
can be manifestly shown by using the transverse and longitudinal 
projection tensors,
$P_{\mu}^{\alpha \perp}=\eta_{\mu}^{\alpha}-q_{\mu}q^{\alpha}/q^2$ and
$P_{\mu}^{\alpha \parallel}=q_{\mu}q^{\alpha}/q^2$:
\beq
\label{eq:definition}
\langle j_{\mu} j_{\nu}^5 \rangle_{\hat F} = \frac{Q^2}{4\pi^2} 
P_{\mu}^{\alpha \perp} \left[P_{\nu}^{\beta \perp} w_T(q^2) 
  + P_{\nu}^{\beta \parallel} w_L(q^2) 
\right] \tilde F_{\alpha \beta},
\eeq
where $Q^2=-q^2$.

The result for $w_L$ is well-known \cite{Adler:1969gk, Bell:1969ts}:
\beq
\label{eq:long}
w_L(Q^2) = \frac{2N_c}{Q^2}.
\eeq 
This quantity does not receive corrections~\cite{Adler:1969er}.
At the level of hadrons, the $1/Q^2$ singularity in Eq.~(\ref{eq:long}) 
is accounted for by the massless pion.

On the other hand, the result for $w_T$ 
is known perturbatively \cite{Vainshtein:2002nv},
\beq
w_T^{\rm pert}(Q^2) = \frac{N_c}{Q^2}.
\eeq
This quantity does not receive perturbative corrections 
as first shown by Vainshtein 
\cite{Vainshtein:2002nv} but it receives nonperturbative corrections 
\cite{Czarnecki:2002nt, Knecht:2003xy}.
In the next section, we will show that the nonperturbative corrections
are given in Eq.~(\ref{eq:result}) for any $Q^2$ in the 
class of holographic QCD models mentioned above.

\section{Holographic description}
\label{sec:holography}
\subsection{Setup}
The five-dimensional (5D) action of the holographic dual of our theory
consists of a Yang-Mills (YM) and a 
Chern-Simons (CS) terms with a $\U(N_f)$ gauge group,
\begin{align}
  S &= S_{\rm YM}+S_{\rm CS} \\
  \label{eq:YM}
  S_{\rm YM} &= -\int\! d^5x \tr \left[-f^2(z){\cal F}_{z\mu}^2
  + \frac{1}{2g^2(z)}{\cal F}_{\mu\nu}^2 \right], \\
  \label{eq:CS}
  S_{\rm CS} &= \kappa \int\! \tr 
  \left[{\cal AF}^2-\frac{i}{2}{\cal A}^3{\cal F}-\frac{1}{10}{\cal A}^5
\right].
\end{align}
Here and below, $z$ is the fifth coordinate
which runs from $-z_0$ to $z_0$ ($z_0>0$); the
Greek indices $\mu, \nu, \cdots$ denote the 4D boundary coordinates and
the Latin indices $M, N, \cdots$ denote the bulk 5D coordinates.
${\cal A}(x,z)={\cal A}_M dx^M$ is the 5D $\U(N_f)$ gauge field and
${\cal F}=d{\cal A}+i{\cal A} \wedge {\cal A}$ is the field strength.
They are decomposed as ${\cal A}= {\cal A}^a t^a$ and 
${\cal F}= {\cal F}^a t^a$.

The functions $f(z)$ and $g(z)$ with the conditions 
$f(-z)=f(z)$ and $g(-z)=g(z)$ 
(required by parity) are related to the metric of the bulk.
For example, in the ``cosh" model considered in \cite{Son:2003et},
$f(z) \sim \cosh(z)$ and $g(z)={\rm const}$ with $z_0=\infty$,
and in the Sakai-Sugimoto model \cite{Sakai:2004cn}, 
$f(z) \sim (1+z^2)^{1/2}$ and $g(z) \sim (1+z^2)^{1/6}$ with $z_0=\infty$.
In order to keep discussion general, we will leave $f(z)$ and $g(z)$ 
unspecified; our results below will be valid for any choice of 
$f(z)$ and $g(z)$
[provided that $\int_{-z_0}^{z_0}\!dz\,f^{-2}(z)$ 
is convergent, see Eq.~(\ref{eq:decay-const})].
On the other hand, $\kappa$ will be fixed as $\kappa=N_c/(24\pi^2)$ 
to reproduce the correct anomaly in QCD [see Eq.~(\ref{eq:kappa})].
In the top-down approach, the CS term with $\kappa=N_c/(24\pi^2)$ 
is obtained from the effective action of the probe 
D8-branes~\cite{Sakai:2004cn}.

As shown in Ref.~\cite{Son:2003et}, this theory can be interpreted as
a theory of mesons, which includes infinite towers of vector mesons
and axial-vector mesons, and one massless pion.  We decompose the
gauge field ${\cal A}(x,z)$ into a parity-even part $V(x,z)$
and a parity-odd part $A(x,z)$, 
\begin{equation}
\begin{split}
  & {\cal A}(x,z) = V(x,z) + A(x,z),\\
   & V(-z)=V(z), \qquad A(-z)=-A(z),
\end{split}
\end{equation}
which 
correspond to vector and axial-vector modes, respectively.  Then
boundary conditions are imposed at $z=0$ (which we call the IR brane):
$V'(0)=0$ and $A(0)=0$, where the derivative is taken with respect to
$z$.  Chiral symmetry is broken due to the different boundary 
conditions of $V$ and $A$.
The boundary conditions at $z=\pm z_0$ (the UV branes) are
the external gauge fields,
\beq
\label{eq:bc}
{\cal A}(z_0)=A_L \equiv V+A, \qquad
{\cal A}(-z_0)=A_R \equiv V-A.
\eeq

Let us first recall the computation of two-point functions of currents
in the absence of the external field $\hat F$.  For this purpose,
the nonlinear CS term in the action can be dropped.  We will work in
the ${\cal A}_z (x,z)=0$ gauge.  The field ${\cal A}_\mu$ satisfies a
linear differential equation, which is easiest to solve in terms of
the Fourier components ${\cal A}(x,z)$.  The solution depends linearly
on the boundary conditions, $V^a_{\mu 0}$ and $A^a_{\mu 0}$, through
the mode functions $V(q,z)$, $A(q,z)$, and $\psi(z)$,
\begin{equation}
  {\cal A}^a_\mu(q,z) =
  V(q,z) P_\mu^{\alpha \perp} V_{\alpha 0}^a(q)
  + A(q,z) P_\mu^{\alpha\perp} A_{\alpha 0}^a(q)
  + P_\mu^{\alpha\parallel} V_{\alpha 0}^a (q) 
  -\psi(z)P_ \mu ^{\alpha \parallel}A_{\alpha 0}^a(q)
\end{equation}
(as will be seen later, the mode function for the longitudinal part of
$V$ is simply 1).  The mode functions satisfy the boundary conditions
\begin{equation}
  V(q,\pm z_0)=1, \qquad A(q,z_0)= -A(q,-z_0) = 1, \qquad
  \psi(z_0) = -\psi(-z_0)=1.
\end{equation}

The linearized field equations are given by
\begin{gather}
\label{eq:eom-trans_V}
  \partial_z \left[ f^2(z) \partial_z V(Q,z) \right] 
    - \frac{Q^2}{g^2(z)} V(Q,z)=0, \\
\label{eq:eom-trans_A}
  \partial_z \left[ f^2(z) \partial_z A(Q,z) \right] 
    - \frac{Q^2}{g^2(z)} A(Q,z)=0,
\\
\label{eq:eom-long}
\partial_z \left[ f^2(z) \partial_z \psi (z) \right] = 0,
\end{gather}
where $Q^2 = -q^2$.  We note that $V$ and $A$ are two linearly
independent solutions to the same differential equation, so their
Wronskian should be independent of $z$:
\begin{equation}\label{Wronskian}
  f^2(z) [V(Q,z)A'(Q,z) - A(Q,z) V'(Q,z)] = W(Q).
\end{equation}

On the other hand, Eq.~(\ref{eq:eom-long}) can be solved as
\beq
\label{eq:pi}
\psi(z)=C_{\pi} \int_0^z\! \frac{dz'}{f^2(z')}, \qquad 
C_{\pi} \int_0^{z_0}\!\frac{dz}{f^2(z)}=1.
\eeq
The longitundal vector mode function satisfies the same equation as
Eq.~(\ref{eq:eom-long}), but with the boundary value of 1 at both $\pm
z_0$.  This function is identically 1.

Using the field equations, one can perform integration in the action
by parts and the integral reduces to the boundary values at $z=\pm z_0$:
\beq
S_{\rm YM}=\frac{1}{2}\left. \int\! \frac{d^4q}{(2\pi)^4} f^2(z) 
{\cal A}_{\mu}^a(q,z)\partial_z{\cal A}^{\mu a}(q,z)\right|_{z=-z_0}^{z=+z_0}.
\eeq
Differentiating the action twice with respect to the boundary value
$V_{\mu 0}^{a}$,
one finds the vector current correlation function,
\begin{gather}
 i\!\int\! d^4 x\, e^{iqx}\langle j_{\mu}^a(x)j_{\nu}^b(0) \rangle
=\delta^{ab}Q^2 P^{\perp}_{\mu \nu}\Pi_V(Q^2),
\\
\label{eq:Pi_V}
\Pi_V(Q^2) = \left. \frac{1}{Q^2}f^2(z) V(Q,z) V'(Q,z)\right|_{z=-z_0}^{z=+z_0}
  = \left. \frac2{Q^2}f^2(z) V'(Q,z)\right|^{z=z_0}
,
\end{gather}
and similarly for $\Pi_A$.
Especially, the pion decay constant $f_{\pi}$ can be obtained from
the longitudinal part of the axial current correlation function, 
\beq
f_{\pi}^2= \left. f^2(z) \psi(z) \psi'(z) 
\right|_{z=-z_0}^{z=+z_0} = 2C_{\pi},
\eeq
or equivalently,
\beq
\label{eq:decay-const}
\frac{4}{f_{\pi}^2}=\int_{-z_0}^{z_0}\!\frac{dz}{f^2(z)}.
\eeq
This expression is consistent with the one obtained in 
\cite{Son:2003et} as it should be. We assume that the right
hand side of Eq.~(\ref{eq:decay-const}) is convergent so that
$f_{\pi}$ is finite.

\subsection{Longitudinal and transverse triangle anomalies}

We then take into account the effect of the CS term induced
by the weak background field $\hat F_{\mu \nu}$.  We will work in the
limit of weak background field $\hat F$, and expand to linear order in
$\hat F$.  For the computation of $w_L$ and $w_T$, we can neglect the
nonlinear terms in the YM action, because they do not include
$VVA$ interactions accompanied with $\epsilon_{\mu \nu \alpha \beta}$
tensor.

First we note that we do not have to find the correction to the
classical solution that comes from the CS action.  
Indeed, our old solution (to the Maxwell equation) 
is an extremum of the classical action, 
and hence a small change in the solution does not change the
YM action to linear order.  All we have to do is to substitute
our old solution into the CS action.

We then note that, unlike the YM action, the CS action
is not gauge-invariant (up to boundaries).
In order for ${\cal A}_z=0$, we carry out the gauge transformation
\beq
{\cal A}_{M} \rightarrow {\cal A}_{M} - \partial_M \Lambda, \\
\Lambda=\int_0^z\!dz\,\frac{f_{\pi}}{2 f^2(z)} \pi(x).
\eeq
This keeps the transverse part of ${\cal A}_M$ unchanged, but 
changes the longitudinal part as 
$\partial_{\mu} A_z \rightarrow -\partial_z A_{\mu}^{\parallel}$.
The contributions to $w_L$ and $w_T$ come from the first term 
in Eq.~(\ref{eq:CS}) after the gauge transformation:
\beq
\label{eq:CS2}
S_{\rm CS} \supset 3\kappa d^{ab} \tilde F_{\mu \nu} 
\int\! d^5x\, (\partial_z V_{\mu}^a A_{\nu}^b - V_{\mu}^a \partial_z A_{\nu}^{b \parallel}).
\eeq
Differentiating $S_{\rm CS}$ with respect to 
$V_{\mu 0}^a$ and $A_{\nu 0}^b$, one obtains $w_L$ and $w_T$.
Remembering the definition (\ref{eq:definition}), one has\footnote{
The expression for $w_L$ is similar to the one obtained in \cite{Gorsky:2009ma}, 
but is different by the boundary value.}
\beq
\label{eq:w_L}
w_L(Q^2)&=& \frac{24\pi^2 \kappa}{Q^2}
\int_{-z_0}^{z_0}\! dz\, \psi'(z) V(0,z) = \frac{48\pi^2 \kappa}{Q^2},
\\
\label{eq:w_T}
w_T(Q^2)&=&\frac{24\pi^2 \kappa}{Q^2}\int_{-z_0}^{z_0}\!dz\, A(Q,z) V'(Q,z),
\eeq
where we took the on-shell amplitude for $w_L$ and used $V(0,z)=1$.
Matching between Eq.~(\ref{eq:w_L}) and the QCD result (\ref{eq:long}) 
leads to the identification:
\beq
\label{eq:kappa}
\kappa=\frac{N_c}{24\pi^2}.
\eeq
As seen from our derivation, $w_L$ is fixed by the
boundary values alone reflecting its topological nature,
whereas evaluating $w_T$ needs dynamical 
information encoded in the field equations.
Performing the integral by parts and using Eq.~(\ref{Wronskian}),
$w_T$ can be written as
\beq
w_T&=&\frac{N_c}{Q^2} 
-\frac{N_c}{2Q^2}\int_{-z_0}^{z_0}\!dz\, (VA'- AV') \nonumber \\
\label{eq:w_T-W}
&=&\frac{N_c}{Q^2} - \frac{N_c}{2Q^2} 
\int_{-z_0}^{z_0}\!dz\, \frac{W(Q)}{f^2(z)},
\eeq
Using the pion decay constant (\ref{eq:decay-const}), 
Eq.~(\ref{eq:w_T-W}) reduces to
\beq
\label{eq:w_T-gravity}
w_T = \frac{N_c}{Q^2} - \frac{2N_c}{f_{\pi}^2 Q^2} W(Q),
\eeq
On the other hand, from Eq.~(\ref{eq:Pi_V}), one obtains
\beq
\label{eq:A-V}
\Pi_A - \Pi_V = \frac 2{Q^2} W(Q).
\eeq
Combining Eqs~(\ref{eq:w_T-gravity}) and (\ref{eq:A-V}),
one finally arrives at the relation 
\beq
\label{eq:trans}
w_T(Q^2)=\frac{N_c}{Q^2}-\frac{N_c}{f_{\pi}^2}
  \left[\Pi_{A}(Q^2) - \Pi_{V}(Q^2)\right],
\eeq
for any $Q^2$.
It is clear from our derivation that this relation holds independently of 
$f(z)$ and $g(z)$ (i.e., the metric of the gravity).
This relation for $w_T$, which leads to the strong constraints 
between the resonance parameters, as we will show below,
may be called the ``anomaly matching for resonances"
as an analogue of $w_L$.

Using the both relations for $w_L$ and $w_T$, 
one also has\footnote{In order to 
obtain Eq.~(\ref{eq:left-right}) from Eqs.~(\ref{eq:w_L})
and (\ref{eq:trans}), one has to add a local counter term proportional to 
$q^2 \tilde F_{\mu \nu}$.}
\beq
\label{eq:left-right}
\langle j_{\mu}^{L} j_{\nu}^{R} \rangle_{\hat F} &=& -\frac{N_c Q^2}{2\pi^2 f_{\pi}^2}
\Pi_{LR}(Q^2) P_{\mu}^{\alpha \perp} P_{\nu}^{\beta \perp}\tilde 
F_{\alpha \beta},
\eeq
for arbitrary $Q^2$, where $j_{\mu}^{La}=\bar q_L \gamma_{\mu} t^a q_L$ is the 
left-handed current and $j_{\mu}^{Ra}=\bar q_R \gamma_{\mu} t^a q_R$
is the right-handed current. The form of this expression except the
proportionality coefficient is fixed solely by the chiral symmetry 
$\SU(N_f)_L \times \SU(N_f)_R$; what we obtained here is the exact 
coefficient $-N_c Q^2/(2\pi^2 f_{\pi}^2)$ 
including the $Q^2$-dependence.

\subsection{Sum rules for resonances}
We shall consider the implications of the relation (\ref{eq:trans}) 
in terms of resonances ($\rho$ meson, $a_1$ meson, and so on).
In the large $N_c$ limit, a tower of resonances with the decay widths 
$\Gamma \sim 1/N_c$ are well defined.
We denote the $i$-th vector meson as $V_i$ ($i=1,2,\cdots$) and
$j$-th axial-vector meson as $A_j$ ($j=1,2,\cdots$).
The wave functions for $V_i$ and $A_j$ in the fifth dimension
$b_{V_i, A_j}(z)$ and their masses $m_{V_i, A_j}$
can be determined by decomposing Eqs.~(\ref{eq:eom-trans_V}) and 
(\ref{eq:eom-trans_A}) into each mode with $q^2=m_{V_i, A_j}^2$,
respectively:\footnote{In our notation, 
$g(z)$ is absorbed into $b_{V_i, A_j}(z)$ compared with the one 
in \cite{Son:2003et}: 
$g(z)b_{V_i, A_j}(z) \rightarrow b_{V_i, A_j}(z)$.}
\beq
\label{eq:eom-mode}
(f^2 b_{V_i}')'=-\frac{m_{V_i}^2}{g^2} b_{V_i}, \qquad
(f^2 b_{A_j}')'=-\frac{m_{A_j}^2}{g^2} b_{A_j}.
\eeq
These functions are subject to the boundary conditions 
$b_{V_i}(-z)=b_{V_i}(z)$, $b_{A_j}(-z)=-b_{A_j}(z)$, 
and $b_{V_i}(\pm z_0)=b_{A_j}(\pm z_0)=0$ with the normalization condition
\beq
\int_{-z_0}^{z_0}\!dz\, \frac{1}{g^2(z)}b_n(z) b_m(z) = \delta_{nm}.
\eeq
The gauge fields $V(Q,z)$ and $A(Q,z)$ can be expanded as
\beq
\label{eq:V-exp}
V(Q,z)&=&
\sum_{i} \frac{g_{V_i}}{Q^2+m_{V_i}^2}b_{V_i}(z), \\
\label{eq:A-exp}
A(Q,z)&=&
\sum_{j} \frac{g_{A_j}}{Q^2+m_{A_j}^2}b_{A_j}(z) - \psi(z).
\eeq
Here $g_{V_i, A_j}$ are the vector and axial-vector meson decay constants defined by 
\beq
\langle 0|j_{\mu}^a(0)|V_i^b(p,\epsilon )\rangle &=& g_{V_i}\delta^{ab}\epsilon_{\mu},
\\
\langle 0|j_{\mu}^{5a}(0)|A_j^b(p,\epsilon )\rangle &=& g_{A_j}\delta^{ab}\epsilon_{\mu},
\eeq
which can be found from Eq.~(\ref{eq:Pi_V}),
\beq
g_{V_i}&= -&\left. f^2(z) b_{V_i}'(z) \right|_{-z_0}^{+z_0}, \\
g_{A_j}&= -&\left. f^2(z) b_{A_j}'(z) \right|_{+z_0}
-\left. f^2(z) b_{A_j}'(z) \right|_{-z_0}.
\eeq
We also define the $\gamma V_i \pi$-couplings $g_{\gamma V_i \pi}$
and the $\gamma V_i A_j$-couplings $g_{\gamma{V_i}{A_j}}$ in 4D QCD:
\beq
{\cal L}_{\gamma{V_i}\pi} &=&
d^{ab} \epsilon^{\mu \nu \alpha \beta} g_{\gamma{V_i}{\pi}}
V_{i\mu}^a \partial_{\nu}\pi^b \partial_{\alpha} \hat V_{\beta}, \\
{\cal L}_{\gamma{V_i}{A_j}} &=&
d^{ab} \epsilon^{\mu \nu \alpha \beta} g_{\gamma{V_i}{A_j}}
V_{i\mu}^a A_{j\nu}^b \partial_{\alpha} \hat V_{\beta}.
\eeq
From Eq.~(\ref{eq:CS2}), these couplings are given by\footnote{
Due to the identity $V(0,z)=\sum_k \frac{g_{V_k}}{m_{V_k}^2}b_{V_k}(z)=1$,
the on-shell photon in the three-point couplings can be replaced 
by the whole tower of vector mesons coupled to the photon 
as a manifestation of the vector meson dominance \cite{Sakai:2004cn}:
\beq
g_{\gamma V_i \pi}&=&\sum_k g_{V_k V_i \pi}\frac{g_{V_k}}{m_{V_k}^2}, \qquad
g_{V_k{V_i}{\pi}} = \frac{N_c}{4\pi^2 f_{\pi}}\int_{-z_0}^{z_0}\! dz\, b_{V_k}(z) b_{V_i}(z) \psi'(z),
\nonumber
\\
g_{\gamma V_i A_j}&=&\sum_k g_{V_k V_i A_j}\frac{g_{V_k}}{m_{V_k}^2}, \qquad
g_{V_k{V_i}{A_j}} = \frac{N_c}{4\pi^2}\int_{-z_0}^{z_0}\! dz\, b_{V_k}(z) b_{V_i}'(z) b_{A_j}(z).
\nonumber
\eeq
The quantities $g_{\gamma V_i \pi}$ and $g_{\gamma V_i A_j}$ will be regarded
as the ``effective three-point couplings" in this respect.}
\beq
g_{\gamma{V_i}{\pi}} &=&
\label{eq:gamma-V-pi}
\frac{N_c}{4\pi^2 f_{\pi}}\int_{-z_0}^{z_0}\! dz\, b_{V_i}(z) \psi'(z).
\\
\label{eq:gamma-V-A}
g_{\gamma{V_i}{A_j}} &=&
\frac{N_c}{4\pi^2}\int_{-z_0}^{z_0}\! dz\, b_{V_i}'(z) b_{A_j}(z).
\eeq

Now we are ready to write $w_L$ and $w_T$ in terms of the resonance parameters.
Substituting the mode expansions (\ref{eq:V-exp}) and (\ref{eq:A-exp}) 
into Eqs.~(\ref{eq:w_L}) and (\ref{eq:w_T}) and performing the
integration over $z$, one obtains
\beq
\label{eq:w_L-exp}
w_L&=&\frac{4\pi^2}{Q^2}\sum_{i} g_{\gamma{V_i}{\pi}} f_{\pi}
\frac{g_{V_i}}{m_{V_i}^2}, \\
\label{eq:w_T-exp}
w_T&=&\frac{4\pi^2}{Q^2}\sum_{i,j}g_{\gamma{V_i}{A_j}}
\frac{g_{V_i}}{Q^2+m_{V_i}^2}\frac{g_{A_j}}{Q^2+m_{A_j}^2}.
\eeq
Therefore, Eq.~(\ref{eq:w_L}) implies the longitudinal sum rule:
\beq
\label{eq:long-sum-rule}
\sum_i \frac{g_{\gamma{V_i}{\pi}} g_{V_i}}{m_{V_i}^2}=\frac{N_c}{2\pi^2 f_{\pi}},
\eeq
and Eq.~(\ref{eq:trans}) leads to the identity:
\beq
\label{eq:resonance}
\sum_{i,j}g_{\gamma{V_i}{A_j}}
\frac{g_{V_i}}{Q^2+m_{V_i}^2}\frac{g_{A_j}}{Q^2+m_{A_j}^2}
=\frac{N_c Q^2}{4\pi^2 f_{\pi}^2}
\sum_{i,j}\left[\frac{g_{V_i}^2}{m_{V_i}^2(Q^2+m_{V_i}^2)}
-\frac{g_{A_j}^2}{m_{A_j}^2(Q^2+m_{A_j}^2)} \right],
\eeq
for arbitrary $Q^2$.
Multiplying both hand sides of this identity 
by $Q^2+m_{V_i}^2$ and then taking $Q^2 \rightarrow -m_{V_i}^2$ limit, 
one obtains a set of transverse sum rules:
\beq
\label{eq:sum-rule1}
\sum_{j}\frac{{g_{\gamma V_i A_j}}g_{A_j}}{m_{A_j}^2-m_{V_i}^2}
=-\frac{N_c}{4\pi^2 f_{\pi}^2}g_{V_i},
\eeq
for $i=1,2,\cdots$. Similarly,
\beq
\label{eq:sum-rule2}
\sum_{i}\frac{{g_{\gamma V_i A_j}}g_{V_i}}{m_{A_j}^2-m_{V_i}^2}
=-\frac{N_c}{4\pi^2 f_{\pi}^2}g_{A_j},
\eeq
for $j=1,2,\cdots$. These sum rules provide stringent constraints 
between the resonance parameters.

\begin{figure}[t]
\begin{center}
\includegraphics[width=10cm]{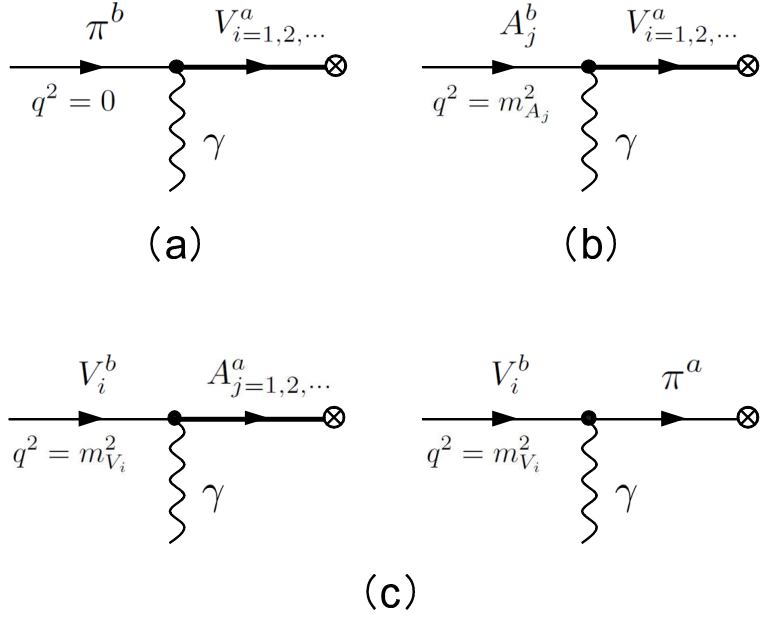}
\end{center}
\vspace{-0.5cm}
\caption{Diagrams contributing to the matrix elements:
(a) $\langle 0|j_{\mu}^a|\pi^b \rangle_{\tilde F}$,
(b) $\langle 0|j_{\mu}^{a}|A_j^b \rangle_{\tilde F}$, and
(c) $\langle 0|j_{\mu}^{5a}|V_i^b \rangle_{\tilde F}$.}
\label{fig:mat-ele}
\end{figure}

These sum rules also fix the matrix elements 
of the vector and axial currents between the vacuum and one
particle state (a pion, a vector meson, or an axial-vector meson) 
in the presence of the soft electromagnetic field 
depicted in Fig.~\ref{fig:mat-ele}.
Substituting Eqs.~(\ref{eq:w_L-exp}) and (\ref{eq:w_T-exp}) 
into the definitions of $w_L$ and $w_T$ in Eq.~(\ref{eq:definition}), 
decomposing them into the sum over $i$ or $j$,
and then using the sum rules, one finds
\beq
\label{eq:mat-ele1}
\langle 0|j_{\mu}^a(0)|\pi^b(q) \rangle_{\tilde F} &=&
i q^{\nu} \frac{N_c}{2\pi^2 f_{\pi}}d^{ab}{\tilde F}_{\mu \nu},
\\
\label{eq:mat-ele2}
\langle 0|j_{\mu}^{a}(0)|A_j^b(q,\epsilon) \rangle_{\tilde F} &=&
-\epsilon^{\alpha}
\left(\eta_{\mu}^{\beta}- \frac{q_{\mu}q^{\beta}}{m_{A_j}^2} \right)
\frac{N_c}{4\pi^2 f_{\pi}^2} g_{A_j}d^{ab}{\tilde F}_{\alpha \beta},
\\
\label{eq:mat-ele3}
\langle 0|j_{\mu}^{5a}(0)|V_i^b(q,\epsilon) \rangle_{\tilde F} &=&
-\epsilon^{\alpha}
\left[\left(\eta_{\mu}^{\beta}- \frac{q_{\mu}q^{\beta}}{m_{V_i}^2} \right)
\frac{N_c}{4\pi^2 f_{\pi}^2} g_{V_i}
-\frac{q_{\mu}q^{\beta}}{m_{V_i}^2}f_{\pi} g_{\gamma V_i \pi} \right]
d^{ab}{\tilde F}_{\alpha \beta}.
\eeq
While Eq.~(\ref{eq:mat-ele1}) will be related to the well-known 
$\pi_0 \rightarrow 2\gamma$ decay if one replaces 
the vector current by an on-shell photon,
Eqs.~(\ref{eq:mat-ele2}) and (\ref{eq:mat-ele3}) are the
new formulas involving resonances.
Remarkably, for fixed isospins $a$ and $b$, the transverse parts of 
the matrix elements (\ref{eq:mat-ele2}) and (\ref{eq:mat-ele3}) are 
respectively proportional to the decay constants $g_{V_i}$ and $g_{A_j}$
with the {\it universal} proportionality coefficient 
independent of species $i$ and $j$ (apart from the transverse projection). 
For example, for $N_f=2$, one has
\beq
\langle 0|j_{\mu}^a|\pi^a \rangle_{\tilde F}^{\parallel} &=&
\tr[{\cal Q}]\frac{N_c}{8\pi^2 f_{\pi}^2}{\tilde F}_{\mu \nu}
\langle 0|j^{\nu 5 a}|\pi^a \rangle,
\\
\langle 0|j_{\mu}^{a}|A_j^a \rangle_{\tilde F}^{\perp} &=&
\tr[{\cal Q}]\frac{N_c}{16\pi^2 f_{\pi}^2}{\tilde F}_{\mu \nu}
\langle 0|j^{\nu 5 a}|A_j^a \rangle, \qquad (j=1,2,\cdots),
\\
\langle 0|j_{\mu}^{5a}|V_i^a \rangle_{\tilde F}^{\perp} &=&
\tr[{\cal Q}]\frac{N_c}{16\pi^2 f_{\pi}^2}{\tilde F}_{\mu \nu}
\langle 0|j^{\nu a}|V_i^a \rangle, \qquad (i=1,2,\cdots),
\eeq
where no summation is taken over $a$. 
We note here that the universality of the proportionality
coefficient originates from the constant value $-N_c/f_{\pi}^2$
with no $Q^2$-dependence in front of the bracket in Eq.~(\ref{eq:trans}).

The above sum rules and resultant matrix elements are generic to any 
theory with a Yang-Mills-Chern-Simons gravity dual in the large $N_c$ limit.
As an example, we explicitly check the sum rules using the ``cosh" model 
\cite{Son:2003et} in Appendix \ref{sec:cosh}.
However, they will not be generally valid in a theory 
incorporating the scalar field corresponding to the chiral condensate 
\cite{Erlich:2005qh, Da Rold:2005zs, Karch:2006pv}
(although we have a different type of sum rules which may be
irrelevant to real QCD).
We provide this counterexample in Appendix \ref{sec:AdS/QCD}.
In the next section, we will discuss that real QCD behaves similarly 
to the former class of theories with the universality 
rather than to the latter counterexample.

If one assumes that sum rules (\ref{eq:sum-rule1}) and (\ref{eq:sum-rule2}) 
are saturated by the lowest resonances $i=j=1$, one has
\beq
\label{eq:prediction1}
g_{V_1}&=&g_{A_1}, \\
\label{eq:prediction2}
g_{\gamma {V_1}{A_1}}&=&-\frac{N_c}{4\pi^2 f_{\pi}^2}(m_{A_1}^2-m_{V_1}^2).
\eeq
Equation~(\ref{eq:prediction1}) is equivalent to the second Weinberg
sum rule $g_{V_1}^2 - g_{A_1}^2=0$ \cite{Weinberg:1967kj}, 
whereas Eq.~(\ref{eq:prediction2}) is a new prediction.
Taking experimental values for these parameters,
we find $g_{\gamma {\rho}{f_1}} \approx -9.2$ 
(and $g_{\gamma {\rho}{a}} \approx -8.0$) for $N_c=3$.\footnote{
A numerical evaluation of (\ref{eq:gamma-V-A}) 
using the specific metric of the Sakai-Sugimoto model 
gives $g_{\gamma {\rho}{f_1}}=-3.8$ \cite{Domokos:2009cq} 
(after matching notation to ours),
which is rather smaller than our prediction
using the truncated sum rules.}
This is not far from the value $|g_{\gamma {\rho}{f_1}}|= 7.6 \pm 1.1$
determined from the experimentally measured decay rate
$\Gamma_{\rm exp}(f_1 \rightarrow \rho^0 + \gamma) 
= 1.34 \pm 0.38$ MeV \cite{PDG:2010} by using
the formula \cite{Kochelev:1999zf}:
\beq
\Gamma(f_1 \rightarrow \rho^0 + \gamma)=
\frac{\alpha d_{30}^2 g_{\gamma \rho f_1}^2}{24} 
\frac{(m_{f_1}^2 + m_{\rho}^2)(m_{f_1}^2-m_{\rho}^2)^3}{m_{\rho}^2 m_{f_1}^5},
\eeq
where $d_{30}=1/4$ for ${\cal Q}={\rm diag}({2}/{3}, -{1}/{3})$.

\section{Real QCD}
\label{sec:QCD}
Let us discuss whether the relation (\ref{eq:trans}) is realized in real QCD.
This is easy to check for $Q^2 \ll \LambdaQCD^2$ where the dynamics 
is governed by the low-lying pions.
Because pions do not contribute to $w_T$, the left hand side of (\ref{eq:trans}) 
should vanish at small $Q^2$. In the right hand side,
pions only contribute to the axial correlator 
$\Pi_A \simeq f_{\pi}^2/Q^2$; the singularities of $1/Q^2$ cancel in total,
and hence, Eq.~(\ref{eq:trans}) is valid.

In the opposite regime, $Q^2 \gg \LambdaQCD^2$, one can make use
of the operator product expansion (OPE) analysis, which is an 
expansion of the correlator in terms of $\LambdaQCD^2/Q^2$.
As usually adopted in the practical applications of the QCD sum rules
\cite{Shifman:1978bx}, we shall neglect the $\alpha_s$-corrections 
and the anomalous dimensions of local composite operators 
in the OPE. Although these simplifications (called the practical OPE) 
are numerically good \cite{Shifman:1998rb}, 
our discussion below is approximate at this level.

\begin{figure}[t]
\begin{center}
\includegraphics[width=10cm]{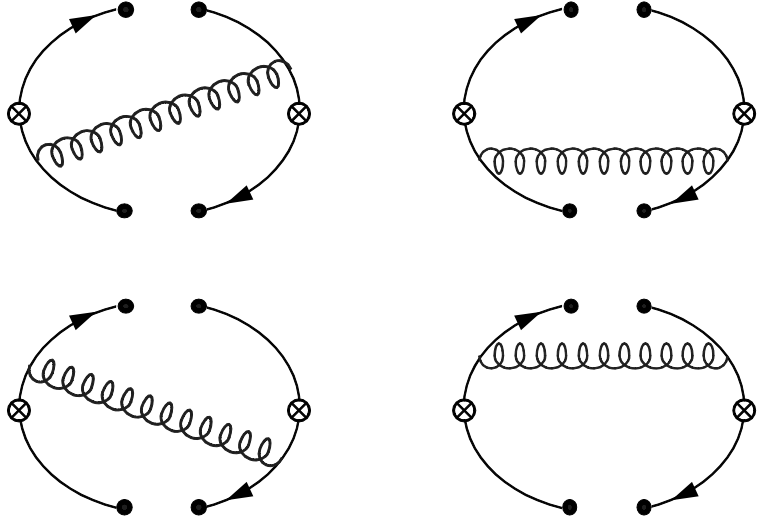}
\end{center}
\vspace{-0.5cm}
\caption{Diagrams contributing to 
$\langle j_{\mu} j_{\nu}^5 \rangle_{\hat F}^{\rm nonpert}$.
The solid and spiral lines denote quarks and gluons respectively.}
\label{fig:4-quark}
\end{figure}

For convenience, look at the relation (\ref{eq:left-right}) instead of (\ref{eq:trans}).
Because of the transformation properties under the $\SU(N_f)_L \times \SU(N_f)_R$ 
symmetry and the Lorentz symmetry,
only the nonperturbative Lorentz pseudo-tensor condensates 
related to chiral symmetry breaking can appear in the OPE of 
$\langle j_{\mu}^{L} j_{\nu}^{R} \rangle_{\hat F}$.
The leading contributions shown in Fig.~\ref{fig:4-quark} 
read \cite{Knecht:2002hr,Czarnecki:2002nt}
\beq
\label{eq:LRV}
\langle j_{\mu}^{L} j_{\nu}^{R} \rangle_{\hat F}
=\frac{1}{2}\langle j_{\mu} j_{\nu}^5 \rangle_{\hat F}^{\rm nonpert}
=-\frac{2g^2}{Q^6}(-q^2 O_{\mu \nu} + q_{\mu}q^{\sigma}O_{\sigma \nu}
-q_{\nu}q^{\sigma}O_{\sigma \mu}),
\eeq
where $g$ is the QCD coupling constant and
$O_{\mu \nu}=\langle (\bar q \gamma_{\mu} \gamma_5 \lambda^a q)
(\bar q \gamma_{\nu} \lambda^a q) \rangle$ is the four-quark condensate 
with the $\SU(N_c)$ color generators $\lambda^a$ ($a=1,2, \cdots, N_c^2-1$).
Using the Fierz transformation
together with the factorization of the four-quark condensate 
(which can be justified in the large $N_c$ limit), one has
\beq
O_{\mu \nu}= - \frac{N_c^2-1}{8N_c^2} \epsilon_{\mu \nu \alpha \beta}
\langle \bar q q \rangle \langle \bar q \sigma^{\alpha \beta} q \rangle.
\eeq 
If we further use the magnetic susceptibility of the chiral condensate $\chi$ 
defined by \cite{Ioffe:1983ju} 
\beq
\langle \bar q \sigma_{\mu \nu} q \rangle=\chi\langle \bar q q \rangle \hat F_{\mu \nu},
\eeq
Equation (\ref{eq:LRV}) reduces to the simple form:
\beq
\label{eq:ope-left}
\langle j_{\mu}^{L} j_{\nu}^{R} \rangle_{\hat F}
=\frac{N_c^2-1}{2N_c^2}\frac{g^2}{Q^4}\chi \langle \bar q q \rangle^2 
P_{\mu}^{\alpha \perp} P_{\nu}^{\beta \perp} \tilde F_{\alpha \beta}.
\eeq
On the other hand, the leading term in the OPE of $\Pi_{LR}$ is \cite{Shifman:1978bx}
\beq
\label{eq:ope-right}
\Pi_{LR}(Q^2)=-\frac{g^2}{Q^6}
\langle (\bar q_L \gamma_{\mu} \lambda^a q_L)(\bar q_R \gamma_{\mu} \lambda^a q_R) \rangle
=\frac{N_c^2-1}{4N_c^2}\frac{g^2}{Q^6}\langle \bar q q \rangle^2.
\eeq

From Eqs.~(\ref{eq:ope-left}) and (\ref{eq:ope-right}), 
that the relation (\ref{eq:left-right}) holds in QCD at large $Q^2$ amounts 
to the condition for $\chi$ to take a special value:
\beq
\label{eq:chi}
\chi=-\frac{N_c}{4\pi^2 f_{\pi}^2}.
\eeq
Interestingly, this is the same value obtained in another way 
assuming the pion dominance in the OPE of $w_L$ 
when one turns on the quark masses \cite{Vainshtein:2002nv}
(see Appendix \ref{sec:chi}).
These results suggest that the relation (\ref{eq:trans})
is valid at least approximately in real QCD at both small and large $Q^2$.

\section{Conclusions}
In this paper, we have shown a relation for the transverse part of 
triangle anomalies (the ``anomaly matching for resonances") 
in holographic QCD. Our relation provides a set of sum rules 
involving the masses, decay constants and couplings between resonances, 
and leads to the formulas for the matrix elements of the vector 
and axial currents in the presence of the soft electromagnetic field.
These results are generic to any theory with a Yang-Mills-Chern-Simons 
gravity dual where chiral symmetry is broken by the boundary conditions.

In real QCD, our relation is also valid at least approximately
when the magnetic susceptibility of the chiral condensate 
takes a special value $\chi=-N_c/(4\pi^2 f_{\pi}^2)$.
The uncertainty of our relation in real QCD should be resolved in the future.
This is relevant to the theoretical estimate of the hadronic electroweak 
contribution concerning $\gamma \gamma^* Z$ triangle diagrams 
to the muon anomalous magnetic moment, 
which can be experimentally determined to high precision 
\cite{Miller:2007kk, Jegerlehner:2009ry}.

There are several open questions. Among others, it is desirable to understand
our relation and resulting formulas for the matrix elements 
in the field theoretical point of view. One can also consider 
its generalization to nonzero temperature and/or nonzero baryon chemical potential.
In relation to heavy ion physics, this
may lead to some possible effects on the ``chiral magnetic effect"
\cite{Kharzeev:2007jp, Fukushima:2008xe} considered to explain the fluctuations 
of charge asymmetry in noncentral collisions.

\acknowledgments
The authors thank M.~A.~Stephanov for discussions and 
A.~Gorsky, M.~A.~Stephanov, and A.~Vainshtein for comments on the manuscript.
N.Y. is supported by JSPS Postdoctoral Fellowships for Research Abroad.
This work is supported, in part, by DOE grant DE-FG02-00ER41132.

\appendix
\section{Summary of results for the ``cosh" model}
\label{sec:cosh}
In this appendix, we explicitly check our formulas in Sec.~\ref{sec:holography}
using the ``cosh" model as an example \cite{Son:2003et}:
\beq
g(z)&=&g_5={\rm const}, \\
f(z)&=&\frac{\Lambda}{g_5}\cosh(z),
\eeq
and $z_0=\infty$.
For completeness, we first review the results obtained in \cite{Son:2003et}.
To match the notation, we assign the integer $n$ to $V_i$ and $A_j$
with $n=2i-1$ for odd $n$ and $n=2j$ for even $n$
(due to the alternate states with the opposite parity).
Then the results in \cite{Son:2003et} are\footnote{Note 
that our boundary conditions (\ref{eq:bc}) are chosen so that
the CS action is introduced in the same way as \cite{Sakai:2004cn},
which are different from ${\cal A}(-z_0)=A_L$ and ${\cal A}(z_0)=A_R$ in 
\cite{Son:2003et}. This entails the change of the sign of $b_n(z)$ 
($n$: even) compared with \cite{Son:2003et}.}
\begin{gather}
b_{n}(z) = (-1)^n g_5 c_n \frac{P_n^1(\tanh z)}{\cosh z}, 
\qquad c_n=\sqrt{\frac{2n+1}{2n(n+1)}},
\\
m_{n}^2=n(n+1)\Lambda^2, 
\\
g_{n}=\sqrt{2n(n+1)(2n+1)}\frac{\Lambda^2}{g_5},
\\
f_{\pi}^2=\frac{2\Lambda^2}{g_5^2},
\\
\frac{1}{g_5^2}=\frac{N_c}{24\pi^2},
\end{gather}
where $P_{n}^1(z)$ are the associated Legendre functions.
We then summarize the new results using the formulas in Sec.~\ref{sec:holography}.
Introducing the variables $y=\tanh z$ and $\nu$ satisfying $\nu(\nu+1)=-Q^2$,
the solutions to the field equations (\ref{eq:eom-trans_V}) and (\ref{eq:eom-trans_A}) are
\beq
V(Q,z)&=&-\frac{\pi}{2}\sec(\nu \pi)\sqrt{1-y^2}
[P_{\nu}^1(y) + P_{\nu}^1(-y)],
\\
A(Q,z)&=&\frac{\pi}{2}\sec(\nu \pi)\sqrt{1-y^2}
[P_{\nu}^1(y) - P_{\nu}^1(-y)],
\eeq
where $\sec t \equiv 1/\cos t$.
The relations (\ref{eq:w_T-gravity}) and (\ref{eq:A-V}) are
\beq
\Pi_A(Q^2)-\Pi_V(Q^2)
&=&\frac{N_c}{12\pi}\sec \left(\frac{\pi}{2}\sqrt{1-
\frac{4Q^2}{\Lambda^2}} \right)
\\
&=&\left\{
\begin{array}{ll}  
\displaystyle{\frac{N_c}{6\pi}} e^{-\pi {Q}/{\Lambda}}, & 
Q^2 \gg \Lambda^2, \stru\\
\displaystyle{\frac{N_c}{12\pi^2}} \displaystyle{\frac{\Lambda^2}{Q^2}}, 
\qquad &Q^2 \ll \Lambda^2,
\end{array}\right.
\eeq
\beq
w_T(Q^2)
&=&\frac{N_c}{Q^2}
\left[1-\pi \frac{Q^2}{\Lambda^2}\sec \left(\frac{\pi}{2}\sqrt{1-
\frac{4Q^2}{\Lambda^2}} \right) \right]
\\
&=&\left\{
\begin{array}{ll} 
\displaystyle{\frac{N_c}{Q^2}}-
\displaystyle{\frac{2\pi N_c}{\Lambda^2}} e^{-\pi {Q}/{\Lambda}}, & 
Q^2 \gg \Lambda^2, \stru\\
\displaystyle{\frac{N_c}{\Lambda^2}}, 
\qquad \qquad \qquad \qquad &Q^2 \ll \Lambda^2,
\end{array}\right.
\eeq
Therefore, the following relation is actually satisfied:
\beq
w_T(Q^2)=\frac{N_c}{Q^2}-\frac{N_c}{f_{\pi}^2}\left[\Pi_{A}(Q^2) - \Pi_{V}(Q^2)\right].
\eeq
For the couplings $g_{\gamma V_i \pi}$ and $g_{\gamma V_i A_j}$, 
which we denote $g_{\gamma n \pi}$ and $g_{\gamma n m}$
with $n=2i-1$ and $m=2j$,
there are the ``neighboring rules":
\begin{gather}
g_{\gamma n \pi}=\frac{4\sqrt{3}}{g_5 f_{\pi}}\delta_{n1},
\\
g_{\gamma n m}=
-6(n+1)\sqrt{\frac{n(n+2)}{(2n+1)(2n+3)}}\delta_{n,m-1}
+6n\sqrt{\frac{(n-1)(n+1)}{(2n-1)(2n+1)}}\delta_{n,m+1}.
\end{gather}
Using the above relations, 
one can easily check the longitudinal and transverse sum rules:
\begin{gather}
\frac{g_{\gamma 1 \pi} g_1}{m_1^2}=\frac{N_c}{2\pi^2 f_{\pi}},
\\
\sum_{m=n\pm 1}\frac{{g_{\gamma n m}}g_{m}}{m_{m}^2-m_{n}^2}
=-\frac{3}{g_5}\sqrt{2n(n+1)(2n+1)}
=-\frac{N_c}{4\pi^2 f_{\pi}^2}g_{n},
\\
\sum_{n=m\pm 1}\frac{{g_{\gamma n m}}g_{n}}{m_{m}^2-m_{n}^2}
=-\frac{3}{g_5}\sqrt{2m(m+1)(2m+1)}
=-\frac{N_c}{4\pi^2 f_{\pi}^2}g_{m}.
\end{gather}

\section{AdS/QCD with the chiral condensate}
\label{sec:AdS/QCD}
One can test whether the relation (\ref{eq:trans}) is realized
in the AdS/QCD incorporating the chiral condensate 
\cite{Erlich:2005qh, Da Rold:2005zs, Karch:2006pv}.
We consider the hard-wall model and follow the notations of \cite{Erlich:2005qh}.
The metric is a slice of anti-de Sitter (AdS) space:
\beq
ds^2=\frac{1}{z^2}(-dz^2 + dx^{\mu}dx_{\mu}), \qquad 0<z\leq z_m.
\eeq
The IR cutoff $z_m$ is responsible for the confinement
and fixes the scale of the $\rho$ meson mass $m_{\rho}$ in this theory.
When we are interested in the physics at large $Q^2$ below,
we can limit ourselves to the region of AdS space close to the boundary
and we can take the $z_m \rightarrow \infty$ limit to simplify the computation.

The action of the theory in the 5D bulk is
\beq
S&=&S_{\rm YM}+S_{\rm CS}, \\
S_{\rm YM}&=&\int\! d^5x\, \sqrt{g} \tr{\left[ |DX|^2 + 3|X|^2- \frac{1}{4g_5^2}(F_L^2 + F_R^2) \right]},
\\
S_{\rm CS}&=& \kappa \int [w_5(A_L)-w_5(A_R)],
\eeq
where $D_{\mu} X = \partial_{\mu} X-iA_{L\mu}X+iX A_{R\mu}$, $A_{L,R}=A_{L,R}^a \tau^a$,
$F_{\mu \nu}=\partial_{\mu}A_{\nu} - \partial_{\nu}A_{\mu} - i[A_{\mu},A_{\nu}]$,
and $w(A)=AF^2-\frac{i}{2}A^3F -\frac{1}{10}A^5$. The coefficient $\kappa$ is fixed
in Eq.~(\ref{eq:kappa}).
The expectation value of the scalar field $X$ is 
determined by the classical solution as
\beq
X_0(z)=\frac{1}{2}m_q z+\frac{1}{2}\sigma z^3.
\eeq
In the following, we consider the chiral limit $m_q=0$. 

We introduce the vector and axial-vector fields $V=(A_L+A_R)/2$ and 
$A=(A_L-A_R)/2$ and we work in the $V_z=A_z=0$ gauge, 
letting $V^{\mu}(q,z)=V(q,z)V_0^{\mu}(q)$
with $V_0$ being the source of the vector current (likewise for $A_{\mu}$).
The linearized equations of motion for the transverse parts
$V_{\perp}(q,z)$ and $A_{\perp}(q,z)$ are
\beq
\label{eq:V_perp}
\left( \frac{V_{\perp}'}{z}\right)'
-\frac{Q^2}{z}V_{\perp}=0,
\\
\label{eq:A_perp}
\left( \frac{A_{\perp}'}{z} \right)'
-\frac{Q^2}{z}A_{\perp}-\frac{g_5^2 v^2}{z^3}A_{\perp}=0,
\eeq
with the boundary conditions $V(Q, \epsilon)=A(Q, \epsilon)=1$
and $V'(Q, z_m)=A'(Q, z_m)=0$.
One can also write down the equation of motion for the longitudinal
part $A_{\parallel}$, but it is irrelevant to our discussion
and is omitted here. 

Equation (\ref{eq:V_perp}) can be solved analytically,
\beq
V_{\perp}(Q,z)= Qz \left[K_1(Qz) + I_1(Qz) \frac{K_0(Qz_m)}{I_0(Qz_m)}\right]
\xrightarrow{z_m \to \infty} Qz K_1(Qz),
\eeq
where $K_n$ and $I_n$ are the modified Bessel functions.
Although Eq.~(\ref{eq:A_perp}) does not allow for an analytical solution generally,
one can solve perturbatively for large $Q^2$,
\beq
A_{\perp}=A_0 + A_1 + \dots,
\eeq
with $A_0(Q,z)=V_{\perp}(Q,z)$. The first correction satisfies
\beq
\label{eq:first-correction}
\partial_x^2 A_1 -\frac{1}{x}\partial_x A_{1} -A_1 = \lambda x^4 A_0,
\eeq
where we define $x \equiv Qz$ and $\lambda \equiv g_5^2 \sigma^2 /Q^6$.
The solution to this equation is given by using the Green's function,
\beq
A_1(x)=\int\! dx'\, G(x,x') \lambda x'^4 A_0(x'),
\eeq
where $G(x,x')$ can be obtained from the
solutions to the homogeneous part of 
Eq.~(\ref{eq:first-correction}),
\beq
f_1(x)=xK_1(x), \qquad f_2(x)=xI_1(x),
\eeq
as
\beq
G(x,x')=-\frac{1}{W[f_1, f_2](x')}[f_1(x)f_2(x') \theta(x-x')
+ f_2(x)f_1(x')\theta(x'-x)],
\eeq
with the Wronskian $W[f_1, f_2](x') \equiv f_1 f_2'-f_1' f_2=x'$.
Using the integral,
\beq
\int_0^{\infty}\!dx'\, x'^5 K_1^2(x')=\frac{8}{5},
\eeq
we find the small $z$ behavior of $A_1$:
\beq
A_1(Q,z)=-\frac{4}{5}(Qz)^2\frac{g_5^2 \sigma^2}{Q^6}.
\eeq
This solution near the boundary is sufficient to evaluate the
correlation functions below which are determined by the boundary values
at $z= \epsilon$ or by the integrals dominated by small $z$ regions.

The derivations of the correlation functions are similar to those in
Sec.~\ref{sec:holography} and we simply denote the resultant expressions here.
The transverse parts of the vector and axial current correlation 
functions are
\beq
\Pi_V(Q^2) &=& \left. -\frac{1}{g_5^2 Q^2}\frac{V_{\perp}'(Q,z)}{z} \right|_{z=\epsilon}, \\
\Pi_A(Q^2) &=& \left. -\frac{1}{g_5^2 Q^2}\frac{A_{\perp}'(Q,z)}{z} \right|_{z=\epsilon},
\eeq
Since $\Pi_A(Q^2) \rightarrow f_{\pi}^2/Q^2$ for $Q^2 \rightarrow 0$, 
the pion decay constant reads
\beq
f_{\pi}^2=\left. -\frac{1}{g_5^2}\frac{A_{\perp}'(0,z)}{z}\right|_{z=\epsilon}.
\eeq
The expressions for $w_L$ and $w_T$ are
\beq
w_L(Q^2)&=& -\frac{2N_c}{Q^2}\int_0^{z_m}\! dz\, A_{\perp}'(0,z) V_{\perp}(0,z) 
\xrightarrow{z_m \to \infty} \frac{2N_c}{Q^2},,
\\
w_T(Q^2)&=& -\frac{2N_c}{Q^2}\int_0^{z_m}\! dz\, A_{\perp}(Q,z) V_{\perp}'(Q,z),
\eeq
where we used $V_{\perp}(0,z)=1$.
The result for $w_L$ is consistent with the anomaly matching condition (\ref{eq:long}).\footnote{If we take
finite $z_m$, however, the nonzero but small value of $\psi(z_m)$ 
at the IR brane slightly breaks the anomaly matching (\ref{eq:long}). 
One may improve this point by adding a surface term at the IR brane
\cite{Grigoryan:2008up}.}

Now we are ready to check the validity of the relation (\ref{eq:trans}) in this theory.
Let us first consider small $Q^2$. 
Using $V(0,z)=1$, one can easily check that $\Pi_V$ and $w_T$ vanish
while $\Pi_A(Q^2) \rightarrow f_{\pi}^2/Q^2$;
thus the relation (\ref{eq:trans}) is valid.

On the other hand, for large $Q^2$, 
we can expand $V_{\perp}$ and $A_{\perp}$ near the boundary,
\beq
V_{\perp}(Q,z)&=&1+\frac{1}{4}(Qz)^2\ln(Q^2 z^2)+\dots,
\\
A_{\perp}(Q,z)&=&1+\frac{1}{4}(Qz)^2\ln(Q^2 z^2)-\frac{4}{5}(Qz)^2 \frac{g_5^2 \sigma^2}{Q^6}
+\dots,
\eeq
which lead to (up to contact terms):
\beq
\label{eq:chi-Pi_V}
\Pi_V(Q^2) &=& -\frac{1}{2g_5^2}\ln Q^2,
\\
\Pi_A(Q^2) &=& -\frac{1}{2g_5^2}\ln Q^2 + \frac{8}{5}\frac{\sigma^2}{Q^6},
\\
\label{eq:chi-w_T}
w_T(Q^2) &=& \frac{N_c}{Q^2} - \frac{32N_c}{5} \frac{g_5^2 \sigma^2}{Q^8},
\eeq
where the integral
\beq
\int_0^{\infty}\!dx\, x^2 K_1(x)=2,
\eeq
is used for evaluating $w_T$.
Matching the leading log behavior in Eq.~(\ref{eq:chi-Pi_V}) with the QCD result:
\beq
\Pi_V (Q^2) = -\frac{N_c}{24\pi^2}\ln Q^2,
\eeq
leads to the identification \cite{Erlich:2005qh}:
\beq
g_5^2=\frac{12\pi^2}{N_c}.
\eeq
Combining the above results, one arrives at
\beq
\label{eq:chi-w_T2}
w_T(Q^2) = \frac{N_c}{Q^2}-\frac{48\pi^2}{Q^2}[\Pi_A(Q^2)-\Pi_V(Q^2)],
\eeq
for large $Q^2$.
Clearly, the nonperturbative correction is different from Eq.~(\ref{eq:trans}) 
and from the behavior in real QCD shown in Sec.~\ref{sec:QCD}: 
the coefficient in front of the bracket is $Q^2$-dependent 
but not a constant $-N_c/f_{\pi}^2$.
This difference originates from the OPE of $w_T$ in 
Eq.~(\ref{eq:chi-w_T}) where the nonperturbative correction is
proportional to $1/Q^8$ rather than $1/(f_{\pi}^2Q^6)$.
This will be due the absence of the field corresponding to the operator 
$\bar q \sigma_{\mu \nu} q$ in this theory which is essential
for the relation (\ref{eq:trans}) to be realized in real QCD 
at large $Q^2$.
One may improve this point by adding the tensor field 
$H_{\mu \nu}$ corresponding to the operator $\bar q \sigma_{\mu \nu} q$
in the theory, although it would still require a fine-tuning of parameters 
to reproduce the quantitatively correct OPE in QCD.

In this case, one can still derive a set of transverse sum rules 
[but different type from Eqs.~(\ref{eq:sum-rule1}) and (\ref{eq:sum-rule2})]
for highly excited resonances using Eq.~(\ref{eq:chi-w_T2}).
Since pions do not contribute to $w_T$, 
we have only to consider the contributions from the vector and 
axial-vector mesons for $w_T$.
Similarly to Eq.~(\ref{eq:resonance}), one obtains a relation:
\beq
\sum_{i,j}g_{\gamma{V_i}{A_j}}
\frac{g_{V_i}}{Q^2+m_{V_i}^2}\frac{g_{A_j}}{Q^2+m_{A_j}^2}
&=& 12 \sum_{i,j}\left[\frac{g_{V_i}^2}{m_{V_i}^2(Q^2+m_{V_i}^2)}
-\frac{g_{A_j}^2}{m_{A_j}^2(Q^2+m_{A_j}^2)} \right]
\nonumber \\
& & -\frac{12 f_{\pi}^2}{Q^2} + \frac{N_c}{4\pi^2},
\eeq
for sufficiently large $Q^2$. This provides a set of sum rules for
highly excited states:
\beq
\label{eq:chi-sum-rule1}
\sum_{j}\frac{{g_{\gamma V_i A_j}}g_{A_j}}{m_{A_j}^2-m_{V_i}^2}
&=& 12 \frac{g_{V_i}}{m_{V_i}^2}, \qquad (i \gg 1),
\nonumber \\
\label{eq:chi-sum-rule2}
\sum_{i}\frac{{g_{\gamma V_i A_j}}g_{V_i}}{m_{A_j}^2-m_{V_i}^2}
&=& 12 \frac{g_{A_j}}{m_{A_j}^2}, \qquad (j \gg 1).
\eeq
They also lead to the relations for the transverse parts of the
matrix elements:
\beq
\label{eq:chi-mat-ele1}
\langle 0|j_{\mu}^{5a}|V_i^b \rangle_{\tilde F}^{\perp} &=&
12\epsilon^{\alpha}
\frac{g_{V_i}}{m_{V_i}^2}
d^{ab}{\tilde F}_{\alpha \beta}, \qquad (i \gg 1),
\\
\label{eq:chi-mat-ele2}
\langle 0|j_{\mu}^{a}|A_j^b \rangle_{\tilde F}^{\perp} &=&
12\epsilon^{\alpha}
\frac{g_{A_j}}{m_{A_j}^2}d^{ab}{\tilde F}_{\alpha \beta},
\qquad (j \gg 1).
\eeq
These matrix elements are proportional not only to the decay constants
$g_{V_i}$ and $g_{A_j}$ but also to $1/m_{V_i}^2$ and $1/m_{A_j}^2$, 
respectively:
there is no universality of the proportionality coefficients unlike 
Eqs.~(\ref{eq:mat-ele2}) and (\ref{eq:mat-ele3}).
This again comes from the $1/Q^2$ behavior in front of the bracket 
in Eq.~(\ref{eq:chi-w_T2}), and is different from real QCD where 
this factor should be approximately replaced by a constant value.
Therefore, we expect that real QCD would have the 
properties (\ref{eq:mat-ele2}) and (\ref{eq:mat-ele3}) rather 
than (\ref{eq:chi-mat-ele1}) and (\ref{eq:chi-mat-ele2}).

\section{Magnetic susceptibility of the chiral condensate}
\label{sec:chi}
In this appendix, we review the derivation of the magnetic susceptibility of 
the chiral condensate by Vainshtein \cite{Vainshtein:2002nv}.
Let us consider the modifications of the longitudinal part of 
the correlator (\ref{eq:definition}) when we turn on the degenerate quark masses $m_q$.
For $Q^2\gg \LambdaQCD^2$, the leading contribution can be found using the OPE,
\beq
\langle j_{\mu} j_{\nu}^5 \rangle_{\hat F}^{\parallel}&=&
Q^2 P_{\mu}^{\alpha \perp}P_{\nu}^{\beta \parallel} \epsilon_{\alpha \beta \rho \sigma}
\left[\frac{N_c}{4\pi^2 Q^2}\hat F^{\rho \sigma}-
\frac{2m_q \langle \bar q \sigma^{\rho \sigma} q \rangle}{Q^4}
+{\cal O}\left(\frac{1}{Q^6} \right) \right]
\nonumber \\
\label{eq:UV}
&=&\frac{Q^2}{4\pi^2}P_{\mu}^{\alpha \perp}P_{\nu}^{\beta \parallel} \tilde F_{\alpha \beta}
\left[\frac{2N_c}{Q^2} - \frac{16\pi^2 \chi m_q \langle \bar q q \rangle}{Q^4} 
+ {\cal O}\left(\frac{1}{Q^6} \right) \right],
\eeq
where we used the definition of $\chi$ in Eq.~(\ref{eq:chi}).
For $Q^2 \ll \LambdaQCD^2$, the pion propagator is replaced by the massive one:
\beq
\langle j_{\mu} j_{\nu}^5 \rangle_{\hat F}^{\parallel}
&=&\frac{Q^2}{4\pi^2}P_{\mu}^{\alpha \perp}P_{\nu}^{\beta \parallel} \tilde F_{\alpha \beta}
\frac{2N_c}{Q^2+m_{\pi}^2} \nonumber \\
\label{eq:IR}
&=&
\frac{Q^2}{4\pi^2}P_{\mu}^{\alpha \perp}P_{\nu}^{\beta \parallel} \tilde F_{\alpha \beta}
\left[\frac{2N_c}{Q^2} - \frac{2N_c m_{\pi}^2}{Q^4} + {\cal O}\left(\frac{1}{Q^6} \right) \right].
\eeq
If one assumes the extrapolation of the $1/Q^4$ term in the bracket in Eq.~(\ref{eq:IR}) 
to large $Q^2$ to be matched against that in Eq.~(\ref{eq:UV}), one finds
\beq
\label{eq:chi-match}
\chi=\frac{N_c m_{\pi}^2}{8\pi^2 m_q \langle \bar q q \rangle}.
\eeq
Using the Gell-Mann--Oakes--Renner relation for pions,
\beq
f_{\pi}^2 m_{\pi}^2 = -2m_q \langle \bar q q \rangle,
\eeq
Equation (\ref{eq:chi-match}) reduces to
\beq
\chi=-\frac{N_c}{4\pi^2 f_{\pi}^2}.
\eeq

\end{document}